\documentclass[10pt]{article}

\usepackage{amsmath}
\usepackage{amssymb}
\usepackage{graphicx}
\usepackage{cite}
\usepackage{color} 

\usepackage{setspace} 
\usepackage[labelfont=bf,labelsep=period,justification=raggedright]{caption}

\makeatletter
\renewcommand{\@biblabel}[1]{\quad#1.}
\makeatother

\date{}

\graphicspath{{figures/}}
\DeclareMathOperator*{\mean}{mean}

\begin{document}

% Title must be 150 characters or less
\begin{flushleft}
{\Large
\textbf{Mapping Topographic Structure in White Matter Pathways with Level Set
Trees}}

% Insert Author names, affiliations and corresponding author email.
Brian P. Kent$^{1}$, 
Alessandro Rinaldo$^{1}$,
Fang-Cheng Yeh$^{2}$,
Timothy Verstynen$^{3,\ast}$
\\
\bf{1} Department of Statistics, Carnegie Mellon University, Pittsburgh, PA, USA
\\
\bf{2} Department of Biomedical Engineering, Carnegie Mellon University, Pittsburgh, PA, USA
\\
\bf{3} Department of Psychology and Center for the Neural Basis of Computation,
Carnegie Mellon University, Pittsburgh, PA, USA
\\
$\ast$ E-mail: timothyv@andrew.cmu.edu
\end{flushleft}

\section*{Abstract}
Fiber tractography on diffusion imaging data offers rich potential for
describing white matter pathways in the human brain, but characterizing the
spatial organization in these large and complex data sets remains a challenge.
We show that level set trees---which provide a concise representation of the
hierarchical mode structure of probability density functions---offer a
statistically-principled framework for visualizing and analyzing topography in
fiber streamlines. Using diffusion spectrum imaging data collected on
neurologically healthy controls (N=30), we mapped white matter pathways from
the cortex into the striatum using a deterministic tractography algorithm that
estimates fiber bundles as dimensionless streamlines. Level set trees were
used for interactive exploration of patterns in the endpoint distributions of
the mapped fiber tracks and an efficient segmentation of the tracks that has
empirical accuracy comparable to standard nonparametric clustering methods. We
show that level set trees can also be generalized to model pseudo-density
functions in order to analyze a broader array of data types, including entire
fiber streamlines. Finally, resampling methods show the reliability of the
level set tree as a descriptive measure of topographic structure, illustrating
its potential as a statistical descriptor in brain imaging analysis. These
results highlight the broad applicability of level set trees for visualizing
and analyzing high-dimensional data like fiber tractography output.

\section*{Introduction}
\label{sec:intro}

Fiber tractography on diffusion weighted imaging (DWI) data can provide a
high-resolution map of the anatomical connections between two brain
areas~\cite{Hagmann2006}. The deterministic variant of fiber tractography
generates a set of simulated fiber streamlines that provide rich information
about the topographic structure of white matter pathways~\cite{Descoteaux2009,
Hagmann2008, Wedeen2008}. This method has been used recently to characterize the
sheet-like layout of large, myelinated pathways~\cite{Wedeen2012}, map the
organization of fiber bundles within the same pathway~\cite{Greenberg2012,
Verstynen2011, Verstynen2012}, identify novel neuroanatomical
patterns~\cite{Makris2009, Wang2012, Catani2012, Catani2013} and quantify the
global structural connectivity between large sets of brain
regions~\cite{Hagmann2008, Jarbo2012}, providing a so-called structural
``connectome" of the human brain (see Van Essen et
al.~(2012)~\cite{VanEssen2012}). The topography and connectivity of the
structural connections identified with fiber tractography have also been shown
to relate directly to corresponding functional connectivity~\cite{Honey2009} and
task-evoked functional dynamics~\cite{Greenberg2012, Pyles2013}, highlighting
the relationship between structure and function in neural systems. Despite these
advances, the lack of descriptive metrics for the spatial topography of white
matter pathways remains a standing problem with structural connectivity analysis
(see Jbabdi et al.~(2013)~\cite{Jbabdi2013}).

Clustering is a popular method for summarizing the spatial structure in white
matter tracks~\cite{Moberts2005, O'Donnell2013}, but clustering is often a
difficult and ill-defined task. Many of the proposed approaches, such as fuzzy
c-means~\cite{Shimony2002, Li2010}, spectral clustering~\cite{O'Donnell2005,
Jonasson2005}, diffusion maps~\cite{Wassermann2008}, local linear
embedding~\cite{Tsai2007}, geometric clustering~\cite{Gerig2004, Corouge2004},
and white matter atlas matching~\cite{Prasad2011a, Xia2005}, assume there is a
single well-defined partition of the data into $K$ separate groups, where $K$ is
presumed known {\it a priori}. However, when the data are noisy or have a high
degree of complexity or spatial heterogeneity, as is often the case in
neuroimaging, it is more appropriate to assume the data have multi-scale
clustering features that can be captured by a hierarchy of nested partitions of
different sizes. These partitions and their hierarchy provide a wealth of
information about the data beyond typical clustering results, unburdening the
practitioner from the need to guess the ``right'' number of clusters, providing a
global summary of the entire data set and offering the ability to select
sub-clusters at different levels of spatial resolution depending on the
scientific problem at hand.

There are many well-established hierarchical clustering methods, some of which
have been applied to the problem of fiber track segmentation~\cite{Ding2003,
Guevara2011, Wassermann2010, Zhang2005}. However, these methods often suffer
from a lack of statistical justification. Single linkage clustering, for
example, is known to be inconsistent in dimensions greater than
one~\cite{Hartigan1981} and suffers from the problem of
``chaining''~\cite{Moberts2005}. The dendrograms that result from agglomerative
hierarchical clustering do not indicate the optimal number of clusters; in order
to obtain clusters the practitioner must specify the number of clusters or a
threshold at which to cut the dendrogram. Furthermore, the dendrograms that
result from these methods are rarely used as statistical descriptors in their
own right.

Several recent fiber clustering analyses proposed more sophisticated methods
that also do not require {\it a priori} knowledge of the number of clusters.
Wasserman and Deriche (2008)~\cite{Wassermann2008a} and Zvitia et
al.~(2008)~\cite{Zvitia2008} use the mean-shift clustering algorithm, which
finds clusters that correspond to the modes of an assumed probability density
function. Brun et al.~(2004) use spectral clustering but avoid choosing a
cluster number by doing recursive binary data partitions~\cite{Brun2004}. Wang
et al.~(2011) use a hierarchical Bayesian mixture model over supervoxels to
estimate white matter segmentation, with the number of clusters chosen
automatically by a Dirichlet process. Different clustering scales are achieved
by defining supervoxels of various sizes~\cite{Wang2011}. Many of these methods
are capable of clustering at multiple data resolutions, but this is typically
not the focus and the multi-scale clustering results are typically not exploited
for futher analysis.

In this article we introduce and apply the principles of high-density
clustering~\cite{Hartigan1975} for complex fiber tractography from a
high-angular resolution form of DWI.  We implement a general procedure called
the {\it level set tree} for accurate estimation of nested subsets of
high-density data points. Like other agglomerative clustering methods, the
output of our procedure is a hierarchy of clusters that can be represented using
a dendrogram. But unlike any other hierarchical clustering method, the
dendrogram obtained by the level set tree procedure has a direct probabilistic
interpretation in terms of underlying probability density function (see next
section for details and background). As a result, level set trees provide a
means to represent and visualize  data arising from complex and high-dimensional
distributions that is statistically accurate in the sense of being a faithful
encoding of the level sets of a {\it bona fide} density function. This property
extends to any sub-tree of a level set tree, so that with our procedure it is
possible to extract subsets of data at multiple resolutions while retaining the
same probabilistic faithfulness, effectively allowing for dynamic and
multi-scale clustering that does not require advance knowledge of the true
number of clusters.

In the context of fiber tractography, we show how the mode hierarchy of a level
set tree can be used interactively to visualize spatial patterns and to cluster
topographically similar fiber streamlines. Unlike most clustering methods that
output a single partition of the data, level set trees encode many different
cluster permutations and act as a scaffold for interactive exploration of
clustering behavior. We also show how uncertainty can be captured on the level
set tree, suggesting the potential for using the tree as a summary statistic of
topographic structure. Taken together, our results demonstrate that level set
trees offer a solution for describing the topographies found in fiber streamline
data sets and provide a fundamentally new way of visualizing and analyzing
complex spatial patterns in fiber tractography data sets.

\section*{Methods}
\label{sec:methods}

\subsection*{Level set trees for densities}
\label{sec:background}
Suppose we observe a collection of points $\mathbb{X}_n = \{x_1,\ldots,x_n\}$ in
$\mathbb{R}^d$ and we want to identify and visualize the spatial organization of
$\mathbb{X}_n$ without specific knowledge about the data generating mechanism
and in particular without any \textit{a priori} information about the number of
clusters. To be concrete, think of $\mathbb{X}_n$ as the endpoints in
$\mathbb{R}^3$ of $n$ fiber tracks, which we hope to describe in a way that
is anatomically meaningful. Clustering is a common approach to this goal,
but clustering is typically an ill-defined task because the concept of a cluster
is vaguely defined. Our level set tree methodology, in contrast, extends the
statistically principled approach to clustering in Hartigan
(1975)~\cite{Hartigan1975}.

Assume the data points are independent draws from an unknown probability
distribution $P$ on $\mathbb{R}^d$ with probability density function (hereafter
pdf) $f$. That is, $f$ is a non-negative function such that the probability of
observing a data point inside a  subset $A \subset \mathbb{R}^d$ can be computed
as
\begin{equation}\label{eqn:probability}
	P(A) = \int_{x \in A} f(x) dx,  
\end{equation} 
where the integral is the Lebesgue integral in
$d$-dimensions.\footnote{Technically, $A$ must be a {\it measurable} subset of
$\mathbb{R}^d$ in order for the integral to be well defined. See Billingsley
(2012)~\cite{Billingsley2012}. Throughout we will implicitly assume that
measurability holds. From a practical standpoint, this is inconsequential.}
From this expression one can see that a set $A$ where $f$ takes on large
values has a high probability of containing many of the sample points. As a
result, points in the sample $\mathbb{X}_n$ are likely clustered inside such a
set, so it is natural to define clusters as regions of high density $f$.

To formalize this intuition, fix a threshold value $\lambda \geq 0$ and let
$L_\lambda(f) = \{x \in \mathbb{R}^d \colon f(x) \geq \lambda\}$ be the upper
level set of $f$, i.e.~the set of points whose density values exceed the level
$\lambda$. Call the set of connected components of $L_\lambda(f)$ the
$\lambda$-clusters of $f$. More generally, a high-density cluster of $f$ is a
$\lambda$-cluster for some $\lambda$, $0 \leq \lambda \leq \max_x f(x)$. Notice
that according to this probabilistic definition, the notion of a cluster depends
on the choice of $\lambda$ and that for a fixed $\lambda$ the corresponding set
of clusters will typically not give a partition of $\{x \colon f(x) \geq 0\}$.
Also, for larger values of $\lambda$ the $\lambda$-clusters define regions where
the ratio of probability content to volume is higher.

A key feature of high-density clusters is the tree property: if $A$ and $B$ are
two high-density clusters, then one is a subset of the other or they are
disjoint. This implies that high-density clusters form a hierarchy---the {\it
level set tree of $f$}---that is indexed by the level values $\lambda$. The tree
property is extremely advantageous for data analysis for a number of reasons.
First, the level set tree can be depicted as a dendrogram, from which the
overall hierarchy of clusters of $f$ can be visualized across all possible
levels of $\lambda$. In fact, one can regard the level $\lambda$ as providing a
clustering resolution of sorts, with lower values of $\lambda$ corresponding to
larger and coarser clusters and higher values to smaller, more sharply defined
clusters. Thus, the dendrogram of level sets of $f$ provides a multi-scale
representation of the clustering characteristics of $f$. As a result, the
practitioner is free to choose the scale and the number of clusters to extract,
depending on the goals of the analysis. Contrast this with most popular
clustering algorithms that implicitly adopt a single-scale approach and demand a
choice of the number of clusters. Another advantage of the tree property is that
it allows one to represent and store the entire set of cluster inclusions
efficiently with a compact data structure that can be easily accessed and
queried (see Table \ref{tab:tree} and its description in the Results section.
Finally, the dendrogram can be used in a direct and interactive manner for
visualizing and extracting the clusters at various levels of the tree and for
exploring the clustering features of a data set. With this approach, one can
select a varying number of clusters at the same or different levels of $\lambda$
without having to re-run the algorithm.

\begin{figure}[!ht]
	\begin{center}
		\includegraphics[width=0.65\textwidth]{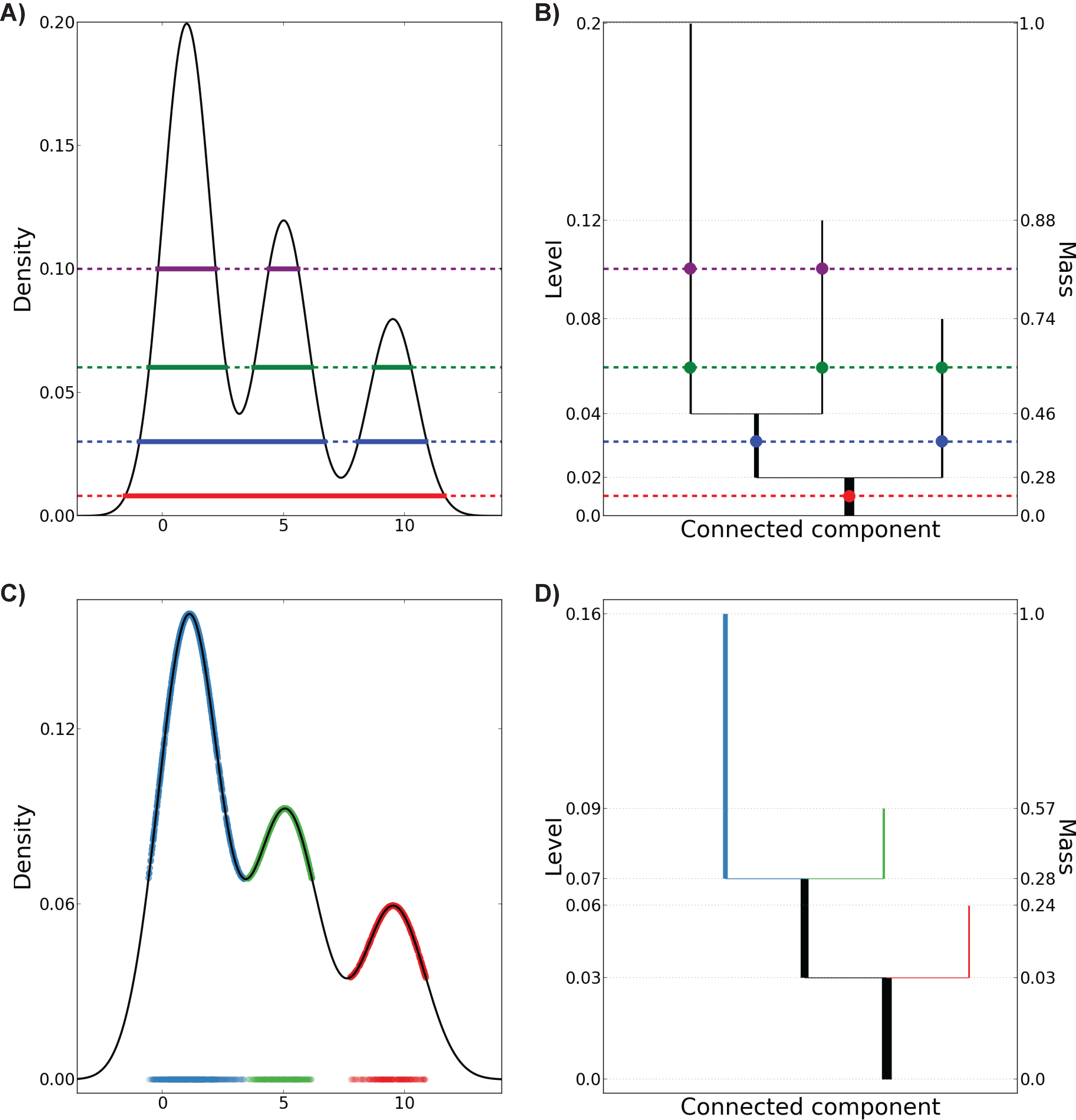}
	\end{center}
	\caption{{\bf Illustration of population and sample level set trees for a
	simple pdf.} A) The true pdf is a mixture of three Gaussians (black curve).
	For each of four example density levels (dotted lines), the high-density
	clusters are indicated by solid line segments. B) Population level set tree
	for the density in panel A. The high-density clusters of panel A are found
	at the intersections of the selected levels (dashed lines) with the tree. C)
	Estimated density (black curve) based on 2,000 data points sampled from the
	pdf in panel A. High-density points belonging to the leaves of the sample
	level set tree in panel D are shown on the horizontal axis and on the
	estimated density function. D) Level set tree estimate based on the sample
	in panel C. Leaves are colored to match corresponding points in the sample.
	For the purpose of illustration the trees in this figure are indexed by
	density levels, while all other trees in this article are plotted on the
	mass scale.}
	\label{fig:figure1}
\end{figure}

Figure \ref{fig:figure1} shows how to read and interpret the level set tree from
a dendrogram. Panel A shows the pdf for a mixture of three Gaussian
distributions and dashed lines representing four values of $\lambda$. For each
level the solid line segment depicts the corresponding clusters. Note that these
are subsets of the real line, even though for illustrative purposes we depict
them at the same level as the corresponding $\lambda$. The tree property can be
seen in the fact that each high-density cluster is a subset of some cluster
portayed immediately below it but is disjoint from all other clusters at the
same level. In panel B the dendrogram of the level set tree is shown; note how
the hierarchy of clusters corresponding to the four levels is respected.
Branching points correspond exactly to levels at which two or more modes of the
pdf, i.e.~new clusters, emerge. Each vertical line segment in this panel
represents the high-density clusters within a single pdf mode. Line segments
that do not branch are considered to be high-density modes, which we call the
leaves of the tree. For simplicity, we tend to treat the terms \emph{dendrogram}
and \emph{level set tree} as synonymous.

\subsection*{Estimating level set trees}
In practice $f$ is not directly observed and one must use the data
$\mathbb{X}_n$ to compute an estimator $\widehat{f}$ of $f$. Under mild
assumptions on $f$ and if the sample size $n$ is large, $\widehat{f}$ is
guaranteed to be very close to $f$ with large probability~\cite{Rao1983} and one
could use the level set tree of $\widehat{f}$ to estimate the level set tree of
$f$ accurately. Unfortunately, computing the $\lambda$-clusters of $\widehat{f}$
is computationally infeasible even in small dimensions because finding the
connected components of the upper level sets of $\widehat{f}$ requires
evaluation of the function on a dense mesh in $\mathbb{R}^d$ and a combinatorial
search over all possible paths connecting any two points of such a mesh.

Instead, we propose a computationally tractable algorithm for level set
clustering that combines and extends procedures outlined originally by Wishart
(1969)~\cite{Wishart1969} and more recently by Maier et
al.~(2009)~\cite{Maier2009}, Chaudhuri and Dasgupta (2010)~\cite{Chaudhuri2010},
and Kpotufe and von Luxburg (2011)~\cite{Kpotufe2011}. At a high level, our
algorithm approximates the level set tree of $\widehat{f}$ by intersecting the
level sets of $\widehat{f}$ with $\mathbb{X}_n$ and then evaluating the
connectivity of each set by graph theoretic means. The details of the method are
given in Tables \ref{tab:lst_algo_main}, \ref{tab:lst_algo_density},
\ref{tab:lst_algo_graph} and \ref{tab:lst_algo_prune}. Our interactive Python
toolbox for level set tree construction, analysis, and clustering is called
\textit{DEnsity-BAsed CLustering (DeBaCl)} and is available at
https://github.com/CoAxLab/DeBaCl.
\begin{table}[!ht]
	\caption{Conceptual level set tree estimation procedure}
	\begin{align*}
		&{\bf Inputs:~} \{x_1,\ldots,x_n\}, k, \gamma \\
		&{\bf Output:~} \widehat{\mathcal{T}} \textrm{, a hierarchy of subsets of } \{x_1,\ldots,x_n\} \\
		&G \leftarrow {\tt ~Compute.knn.graph}(\{x_1,\ldots,x_n\}, k) \\
		&{\bf for~} j \leftarrow 1 {\bf~to~} n: \\
		&\quad \lambda_j \leftarrow {\tt ~Compute.knn.density}(\{x_1,\ldots,x_n\},j, k) \\
		&\quad L_{\lambda_j} \leftarrow \{x_i: \widehat{f}(x_i) \geq \lambda_j\}\\
		&\quad G_j \leftarrow \textrm{~subgraph of~} G \textrm{~induced by~} L_{\lambda_J} \\
		&\quad \textrm{Find the connected components of~} G_{\lambda_j} \\
		&\widehat{\mathcal{T}} \leftarrow \textrm{~dendrogram of connected components of graphs~} G_1,\ldots,G_n, \textrm{~ordered by inclusions}\\
		&\widehat{\mathcal{T}} \leftarrow {\tt ~Prune.tree}(\widehat{\mathcal{T}}, \gamma)\\
		&{\bf Return~} \widehat{\mathcal{T}}
	\end{align*}
	\label{tab:lst_algo_main}
\end{table}

\begin{table}[!ht]
	\caption{{\tt Compute.knn.density}, a procedure for computing the k-nearest neighbor density estimate at the sample points.}
	\begin{align*}
		&{\bf Inputs:~} \{x_1,\ldots,x_n\}, j, k \\
		&{\bf Output:~} \widehat{f}(x_j) \textrm{~a knn density estimate for sample point~} x_j \\
		&r_k^d(x_j) \leftarrow \textrm{k-nearest neighborhood of~} x_j \textrm{~among remaining sample points} \\
		&\widehat{f}(x_j) \leftarrow \frac{k}{n \cdot v_d \cdot r^d_k(x_j)}, \textrm{~where~} v_d \textrm{~is the volume of the Euclidean unit ball in~} \mathbb{R}^d \\
		&{\bf Return~} \widehat{f}(x_j)
	\end{align*}
	\label{tab:lst_algo_density}
\end{table}

\begin{table}[!ht]
	\caption{{\tt Compute.knn.graph}, a procedure for constructing a k-nearest neighbor similarity graph.}
	\begin{align*}
		&{\bf Inputs:~} \{x_1,\ldots,x_n\}, k \\
		&{\bf Output:~} \textrm{k-nearest-neighborhood graph~} G \\
		&{\bf for~} i \leftarrow 1 {\bf~to~} n: \\
		&\quad r_k^d(x_i) \leftarrow  \textrm{~k-nearest neighborhood of~} x_i \textrm{~among the remaining sample points} \\
		&E \leftarrow \emptyset \\
		&{\bf For All~} i < j: \\
		&\quad {\bf If~} \| x_i - x_j \| \leq \max\{r^d_x(x_i),r^d_k(x_j)\}: \\
		&\qquad E \leftarrow E \cup \{ (i,j) \} \\
		&G \leftarrow (\{1,\ldots,n\}, E) \\
		&{\bf Return~} G
	\end{align*}
	\label{tab:lst_algo_graph}
\end{table}

\begin{table}[!ht]
	\caption{{\tt Prune.tree}. Removes small leaf nodes from a level set tree}.
	\begin{align*}
		&{\bf Inputs:~} \widehat{\mathcal{T}} \textrm{~(a hierarchy of subsets of~} \{x_1,\ldots,x_n\}), \gamma \\
		&{\bf Output:~} \textrm{A pruned tree~} \widehat{\mathcal{T}} \\
		&\textbf{For Each~} A \in \widehat{\mathcal{T}}: \\
		&\quad {\bf If~} |A| < \gamma n: \\
		&\qquad \widehat{\mathcal{T}} \leftarrow \widehat{\mathcal{T}} \setminus A \\
		&{\bf Return~} \widehat{\mathcal{T}}
	\end{align*}
	\label{tab:lst_algo_prune}
\end{table}

The first step of our algorithm is to compute a k-nearest neighbor (knn)
similarity graph $G$ with nodes corresponding to $\mathbb{X}_n$ and edges
connecting vertex pairs if either node is one of the $k$ closest neighbors to
the other. In the second step, we compute a knn density estimator (Table
\ref{tab:lst_algo_density})~\cite{Devroye1977,Maier2009}, which we evaluate only
at the $n$ sample points. The parameter $k$ controls the smoothness of the
density $\widehat{f}$; larger values of $k$ produce smoother and flatter density
estimates with small variances but large biases. As a result, choosing a large
$k$ reduces the chance of finding spurious clusters but makes it harder to
detect and separate true clusters that are very close to each other. Choosing a
small $k$ yields nearly unbiased density estimates with large variances. Based
on our experiments and theoretical results (\!\!\cite{Rinaldo2010}
and~\cite{Rinaldo2012}), we tend to favor larger values of $k$.

Construction of the level set tree proceeds by ordering the estimated sample
densities from smallest to largest and iterating over these values. For each
value $\lambda$ in this list, the upper level set is:
\begin{equation}
	L_\lambda = \{x_i: \widehat{f}(x_i) \geq \lambda \}.
\end{equation}
In each iteration we construct an upper level similarity graph $G_\lambda$ by
removing the vertices from $G$ whose sample points are not in $L_\lambda$ and
finding the connected components of $G_\lambda$.

The level set tree is the compilation of connected components over all values of
$\lambda$. The final step is to prune small components of the tree that occur
due to sampling variability or insufficient statistical power. Pruning merges
components that contain fewer than $\gamma n$ data points into other nearby
components. Larger values of $\gamma$ correspond to more aggressive pruning,
where only connected components of large relative size are deemed as separate
clusters. On the other hand, setting $\gamma$ to be very small enhances the
resolution of the clustering procedure but increases the chance of seeing
spurious clusters.

\subsection*{$\alpha$-indexing}
We have defined level set trees based on density thresholds, i.e. values of
$\lambda$. Because this indexing is highly dependent on the height of $f$ (or
$\widehat{f}$), it lacks interpretability (for instance, it is not clear if
$\lambda = 1$ would be a threshold for high or low density regions). To remove
the scale dependence, we instead consider indexing based on probability content
rather than density height. Specifically, let $\alpha$ be a number between $0$
and $1$ and define
\begin{equation}
	\lambda_\alpha = \sup \{ \lambda \colon \int_{x \in L_\lambda(f)} f(x) dx
	\geq \alpha \}
\end{equation}
to be the value of $\lambda$ for which the upper level set of $f$ has
probability content no smaller than $\alpha$~\cite{Rinaldo2012}. The map $\alpha
\mapsto \lambda_\alpha$ gives a monotonically decreasing one-to-one
correspondence between values of $\alpha$ in $[0,1]$ and values of $\lambda$ in
$[0,\max_x f(x)]$. In particular, $\lambda_1 = 0$ and $\lambda_0 = \max_x f(x)$.
Because this map is monotonic, we can express the height of the tree in terms of
the probability content $\alpha$ instead of $\lambda$ without changing the
topology (i.e. number and ordering of the branches) of the tree. The
$\alpha$-indexing is not, however, generally a linear re-indexing of $\alpha$,
so the re-indexed tree will be a deformation of the original tree in which some
branches are dilated and others are compressed. We refer to this probability-
based scale as $\alpha$- or mass-indexing.

To estimate an $\alpha$-indexed tree, we index the level sets of $\widehat{f}
\cap \mathbb{X}_n$ in a similar way. Specifically, for any $\alpha \in \{0,
\frac{1}{n}, \frac{2}{n}, \ldots, \frac{n-1}{n}, 1\}$, we set $\lambda_\alpha$
to be the $\alpha$-quantile of the $n$ estimated sample densities. The
associated hierarchy of subsets $L_{\lambda_\alpha}(\widehat{f}) \cap
\mathbb{X}_n$ is computed as $\alpha$ varies from $1$ to $0$.

We regard $\alpha$-indexing as more interpretable and useful for several
reasons. The $\alpha$ level of the tree indexes clusters corresponding to the $1
- \alpha$ fraction of ``most clusterable" data points; in particular, smaller
$\alpha$ values yield more compact and well-separated clusters. The mass index
can be used for de-noising and outlier removal: to eliminate 5\% of the data
with lowest estimated density, retrieve all the points in the clusters indexed
by levels $\alpha = 0.05$. Scaling by probability content also enables
comparisons of level set trees arising from data sets drawn from different pdfs,
possibly in spaces of different dimensions. The $\alpha$ index is more effective
than the $\lambda$ index for representing regions of large probability content
but low density and is less affected by small fluctuations in density estimates.

\subsection*{Pseudo-density analysis}
\label{ssec:pseudo-density}

A fiber track can be thought of as a set of points sampled along a random curve
in three dimensions. Although probability distributions for these random
functions are well-defined, they cannot be represented with
pdfs~\cite{Billingsley2012}. We can extend level set trees to work with this
type of non-Euclidean data by pseudo-density functions in place of
pdfs~\cite{Ferraty2006}. Pseudo-densities cannot be used to compute
probabilities as in Equation~\ref{eqn:probability}, but they can be regarded as
measures of similarity among points and of the overall connectivity of a space.

To compute the sample level set tree for a collection of fiber tracks, we use
the knn density estimate as in Table \ref{tab:lst_algo_density} but replace the
Euclidean distance with a distance relevant to fibers, expunge the term $v^d$ in
the knn density calculation, and set $d$ arbitrarily to 1. In general this does
not yield a \emph{bona fide} density function, but it is sufficient to induce an
ordering on the data points based on each point's proximity to its neighbors.

We measure the proximity of a pair of fibers with with max-average-min
distance~\cite{Zhang2008}, computed using the Dipy Python module's
bundles\_distances\_mam function~\cite{Garyfallidis2011}. Suppose a set of fiber
tracks $Z_1,\ldots,Z_n$, where $Z_u$ is a sequence of $r$ points
$\{Z_u\}_{i=1}^r$, $Z_{ui} \in \mathbb{R}^3$. The distance between two fibers
$Z_u$ and $Z_w$ is:
 \begin{equation}
 	D(Z_u, Z_w) = \max \bigg\{\mean_i \{ \min_j d(Z_{ui}, Z_{wj})\}, \mean_j \{
	\min_i d(Z_{ui}, Z_{wj})\} \bigg\}
\end{equation}
where $d(Z_{ui}, Z_{wj})$ is the Euclidean distance between the $i$'th point in
fiber $Z_u$ and the $j$'th point of fiber $Z_w$. In practice, points with a
small minimum distance to the other fiber are removed from the computation.
Intuitively this distance matches each point in fiber $Z_u$ to the closest point
in fiber $Z_w$ and \textit{vice versa}, then averages the matched point
distances that are sufficiently large.

Once the distance is computed for each pair of fibers, the pseudo-density
function is evaluated for each fiber and a similarity graph is constructed.
Level set tree construction then follows the procedure in Algorithm
\ref{tab:lst_algo_main}.

\subsection*{Benchmark simulations}
\label{ssec:simulations}
We compared the performance of level set trees in a traditional clustering task
against several popular methods: k-means++~\cite{Arthur2007}, gaussian
mixtures~\cite{Hastie2009}, hierarchical agglomeration with the Ward
criterion~\cite{Hartigan1975}, hierarchical agglomeration with the single
linkage criterion~\cite{Hastie2009}, spectral~\cite{Luxburg2006}, diffusion
map~\cite{Coifman2006}, and DBSCAN~\cite{Ester1996}. Each method was given the
true number of clusters $K$ in order to isolate the effectiveness of the
algorithms from the heuristics for choosing $K$. For the sake of comparison we
used the \textit{fixed $K$} clustering option with level set trees, even though
this ignores the ability of level set trees to automatically choose $K$.

Each method was tested in several three-dimensional data simulations with
varying degrees of realism. The easiest simulation was a mixture of six Gaussian
distributions, the moderate simulation was a mixture of three Gaussian
distributoins and three noisy arcs, and the difficult simulation was a
resampling from a set of 10,000 striatal white matter fiber track endpoints for
a real subject. For the latter scenario, the true clusters were determined by a
careful application of level set tree clustering. To further vary the degree of
difficulty of the clustering tasks, the group means in each scenario were
contracted toward the grand mean by a coefficient $r$, which took 20 values on a
grid ranging from 0.1 to 1.2. Finally, for each simulation type and separation
coefficient, we drew 20 data sets of 5,000 points each. See Figures
5A, C, and E for examples of the simulation scenarios.

Both types of agglomerative hierarchical clustering were implemented with the
\texttt{R hclust} function~\cite{RCoreTeam2012}. K-means++, Gaussian mixture
modeling (GMM), and DBSCAN were implemented with the python module \texttt
{scikit-learn}~\cite{Pedregosa2011}. For DBSCAN we set the neighborhood
parameter $\epsilon$ to be the second percentile of all pairwise distances and
the level set parameter (i.e. the number of neighbors required for a point to be
a core point) to be the first percentile of pairwise distances. Note that DBSCAN
does not allow $K$ to be specified, making it difficult to compare to other
methods.

We used our own implementions for spectral clustering and diffusion maps. For
spectral clustering we constructed a symmetric knn graph on the data, with $k$
set to one percent of the sample size. The points in the first percentile of
degree in this graph were removed as outliers. For diffusion maps we used a
complete similarity graph with Gaussian edge weights:
\begin{equation}
	e(x_i, x_j) = \exp{\left(-\frac{\|x_i-x_j\|^2}{\sigma}\right)}
\end{equation}
with $\sigma$ set to twice the squared median of all pairwise
distances~\cite{Richards2009a}. For both spectral and diffusion map clustering
we use the random walk form of normalized graph Laplacian:
\begin{equation}
	L = D^{-1}(D - W)
\end{equation}
where $W$ is the similarity graph adjacency matrix, and $D$ is the diagonal
degree matrix for $W$~\cite{Luxburg2006}. For diffusion maps the $i$'th
eigenvector $\psi_i$ is scaled by a function of its corresponding eigenvalue
$\rho_i$:
\begin{equation}
	\psi_i' = \left( \frac{1 - \rho_i}{\rho_i} \right) \psi_i
\end{equation}
which creates a multi-scale diffusion map~\cite{Richards2009b}. For spectral
clustering and diffusion maps we use k-means++ to cluster the data after it is
projected into the eigenspace, and for spectral clustering we use a knn
classifier to assign outliers to clusters.

\subsection*{Participants}
\label{ssec:subjects}
Twenty male and ten female subjects were recruited from the local Pittsburgh
community and the Army Research Laboratory in Aberdeen, Maryland. All subjects
were neurologically healthy, with no history of either head trauma or
neurological or psychiatric illness. Subject ages ranged from 21 to 45 years of
age at the time of scanning and four were left handed (2 male, 2 female). All
procedures were approved by the local institutional review board at Carnegie
Mellon University.

\subsection*{Imaging acquisition}
\label{ssec:imaging}
All thirty participants were scanned on a Siemens Verio 3T system in the
Scientific Imaging and Brain Research (SIBR) Center at Carnegie Mellon
University using a 32-channel head coil. We collected a 50 min, 257-direction
DSI scan using a twice-refocused spin-echo EPI sequence and multiple q values
(TR = 9,916 ms, TE = 157 ms, voxel size = $2.4 \times 2.4 \times 2.4$ mm, FoV =
$231 \times 231$ mm, b-max = 5,000 s/mm\textsuperscript{2}, 51 slices). Head-
movement was minimized during the image acquisition through padding and all
subjects were confirmed to have minimal head movement during the scan prior to
inclusion in the template. Since head motion cannot be reliably accounted for
with a scan of this length, we opted instead to minimize head motion during
acquisition.

\subsection*{Diffusion MRI reconstruction}
\label{ssec:dsi}
All DSI images were processed using a q-space diffeomorphic reconstruction
method ~\cite{Yeh2011}, implemented in DSI Studio
(http://dsi-studio.labsolver.org). The coregistration was conducted using a
non-linear spatial normalization approach~\cite{Ashburner1999}, and a total of
16 iterations were used to obtain the spatial mapping function. From here
orientation distribution functions (ODFs) were reconstructed to spatial
resolution of $2 \times 2 \times 2$ mm and a diffusion sampling length ratio of
1.25. To determine the average tractography space, we generated a template image
(the CMU-30 Template) composed of the average whole-brain ODF maps across all 30
subjects.

\subsection*{Fiber tractography}
\label{ssec:tractography}
All fiber tracking was performed using DSI Studio. We used an ODF-streamlined
region of interest (ROI) based approach~\cite{Yeh2010a} similar to that used in
previous studies~\cite{Verstynen2011, Verstynen2012}. Tracks were generated
using an ODF-streamline version of the FACT algorithm~\cite{Basser2000,
Lazar2003, Yeh2010a}. For our initial test-set analysis, in MNI-space, we mapped
two cortico-striatal pathways: lateral frontal (middle frontal gyrus to
striatum) and orbitofrontal (gyrus rectus to striatum). For tractography
analysis on the 30-subject template brain, a whole-brain seeding was used in the
tractography process, with 300 seeds per voxel in the whole-brain mask
(31,100,100 total). For the fiber endpoint analysis and the test-retest
analysis, we only collected 10,000 streamlines per pathway per subject. This was
done to minimize processing and computational demands in the level set tree
generation process and to make equivalent comparisons across pathways with the
same number of samples.

Fiber progression continued with a step size of 1 mm, minimum fiber length of 10
mm, and maximum of 70 mm. To smooth each track, the next directional estimate of
each voxel was weighted by 20 percent of the previous moving direction and 80
percent by the incoming direction of the fiber. The tracking was terminated when
the relative quantitative anisotropy (QA) for the incoming direction dropped
below a preset threshold of 0.2 or exceeded a turning angle of $75^\circ$.

\section*{Results}
\label{sec:results}

\subsection*{Visualizing data with level set trees}
\label{ssec:lst-viz}
Table \ref{tab:tree} shows the raw information in an example level set tree. The
tree is a collection of nodes; each node has start and end $\lambda$ and
$\alpha$ levels, a parent, children (possibly an empty set), and constituent
data points at the node's start level. This information is conveyed more
effectively through a plot of the dendrogram. Figure 1C illustrates a density
estimate for 2,000 points sampled from a mixture of three Gaussian distributions
and Figure 1D shows the estimated level set tree for the sample. Each vertical
line segment of the tree represents the clusters contained in one mode of the
estimated pdf; all of these clusters are subsets of the cluster at the start
level of the mode.
\begin{table}[!ht]
	\caption{\bf{Estimated level set tree information for a simple data
	simulation.}}
	\begin{center}
	\begin{tabular}{c*{7}r}
	\hline
	Node & Start Level & End Level & Start Mass & End Mass & Size & Parent &
	Children\\
	\hline
	0 & 0.000 & 0.005 & 0.000 & 0.021 & 2001 & None & [1, 2]\\
	1 & 0.005 & 0.061 & 0.021 & 0.528 & 1309 & 0 & [3, 4]\\
	2 & 0.005 & 0.165 & 0.021 & 0.998 & 649 & 0 & []\\
	3 & 0.061 & 0.167 & 0.528 & 0.999 & 359 & 1 & []\\
	4 & 0.061 & 0.172 & 0.528 & 0.999 & 295 & 1 & []\\
	\hline
	\end{tabular}
	\end{center}
	\label{tab:tree}
\end{table}

The tree visualization contains several other pieces of information. The height
of each tree node indicates the prominence of the corresponding density mode.
Nodes are sorted so that density modes containing more sample points (i.e. mass)
appear to the left of smaller siblings. The mass of each node is also
proportional to the thickness of its line segment and the amount of whitespace
surrounding its line segment. For example, in Figure 1D, the first
split yields two nodes containing approximately 75\% (black node) and 25\% (red
node) of the mass respectively, so the black segment is thicker and surrounded
by whitespace occupying about 75\% of the width of the plot.

The mode hierarchy shown in a level set tree is a natural platform for
interactively exploring interesting subsets of complicated data; by selecting a
tree branch one can zoom in on structurally coherent groups, ameliorating
overplotting problems to a great extent. Figures 2 and 3 illustrate the use of
level set trees for interactive data visualization on a set of end point
locations from 10,000 streamlines tracked from the lateral frontal cortex to the
striatum. Figure 2B shows each streamline endpoint, color coded by its local
density (higher densities are shown in warmer colors). The tree for this data
set (Figure 2C) is more complicated than the tree in Figure 1. It shows there
are two primary clusters, each of which is further separated into well-defined
sub-groups.
\begin{figure}[!ht]
	\begin{center}
		\includegraphics[width=0.95\textwidth]{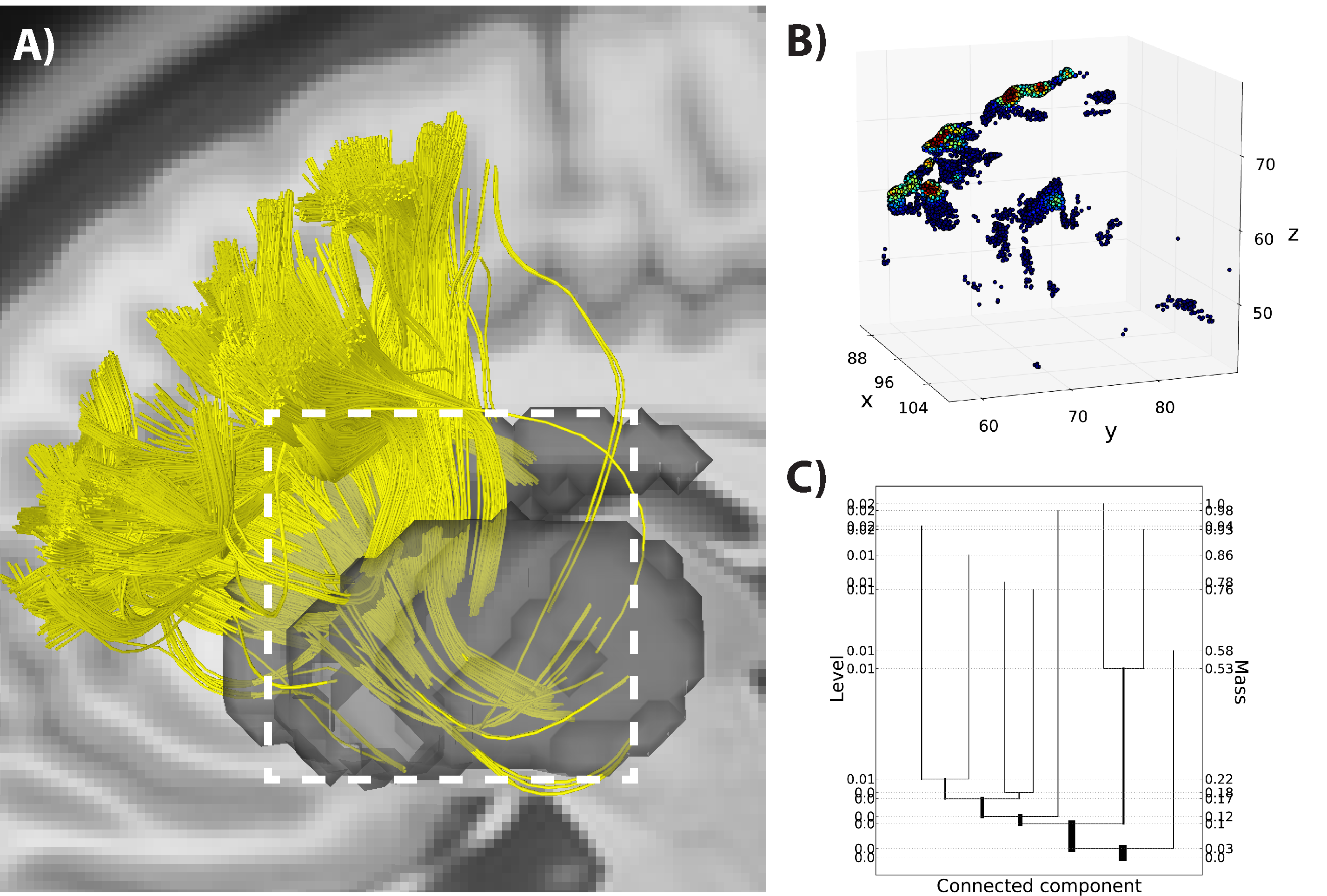}
	\end{center}
	\caption{{\bf Level set tree for cortico-striatal fiber endpoint locations.} A)
	10,000 streamlines (yellow) mapped from the lateral frontal cortex (middle
	frontal gyrus) to the striatal nuclei (caudate nucleus, putamen and nucleus
	accumbens) shown as a gray region of interest (ROI). Data taken from a
	representative subject. B) Endpoint locations (in millimeters) of the
	streamlines shown in panel A, colored by estimated density (red is high). C)
	The corresponding level set tree, which indicates a complex cluster
	structure in these data. A major split occurs when 10\% of the data are
	excluded from the density upper level set, and each branch of the split has
	relevant sub-clusters at various resolutions. Note the lack of information
	in the density level index on this plot, which is a typical outcome.}
	\label{fig:figure2}
\end{figure}

In Figure 3 we use the tree to navigate through this data set. Selecting the
data points associated with one of the large primary branches (Figure 3A and 3B)
shows that this high density region is spatially isolated in a single cluster in
the dorsal portion of the striatum, specifically the dorsal cudate nucleus. By
zooming in on some of the smaller components of the other primary branch (Figure
3C and 3D) we see that these are reflected as independent sub-clusters from the
first branch, with endpoints in the anterior aspect of the caudate near the
shell region of the nucleus, with local density hierarchies within the cluster
(Figure 3D). This illustrates how, by interacting with the different branches of
the level set tree, it is possible to characterize local topographic structures
at different resolutions that reflect known, anatomically distinct sub-regions
of the projections into the caudate~\cite{Haber2010}.
\begin{figure}[!ht]
	\begin{center}
		\includegraphics[width=0.65\textwidth]{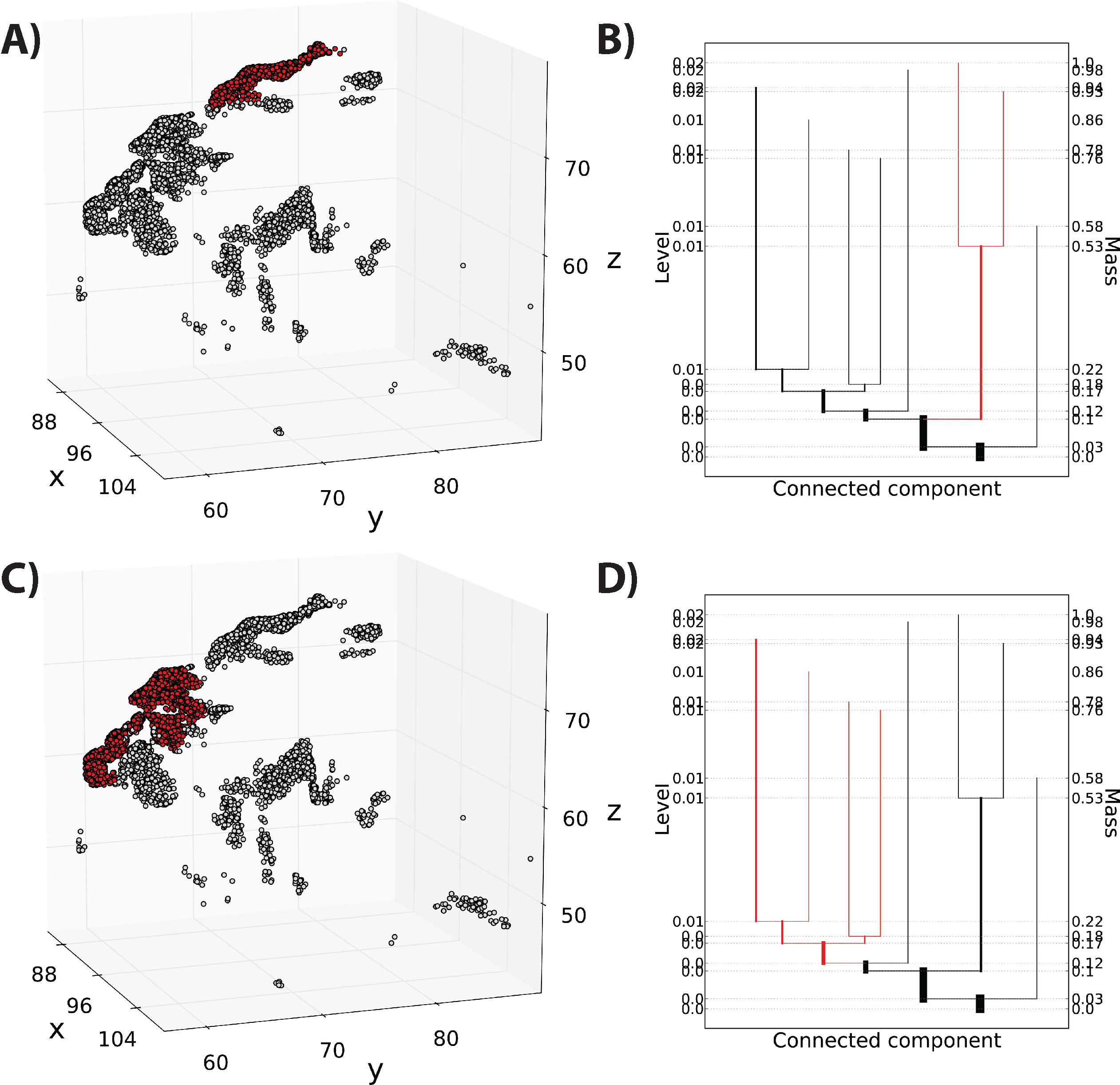}
	\end{center}
	\caption{{\bf Exploring data subsets with a level set tree.} A) Striatal
	endpoints from Figure 2. Red points are members of a selected node of the
	level set tree, shown in red in panel B. C) Striatal endpoints belonging to
	a different mode of the level set tree, shown in panel D.}
	\label{fig:figure3}
\end{figure}

\subsection*{Clustering with level set trees}
\label{ssec:lst-clusters}
Level set trees have several useful properties for solving practical clustering
problems. Most notably, they provide several different ways to obtain cluster
labels, some of which do not require \textit{a priori} knowledge of the number
of clusters. They also remove outliers automatically and allow an investigator
to visualize many different clustering permutations simultaneously and
interactively. Figure 4 shows the output from three of the tree-based cluster
labeling methods applied to the same endpoint distribution shown in Figures 2
and 3 (see Methods).
\begin{figure}[!ht]
	\begin{center}
		\includegraphics[width=0.95\textwidth]{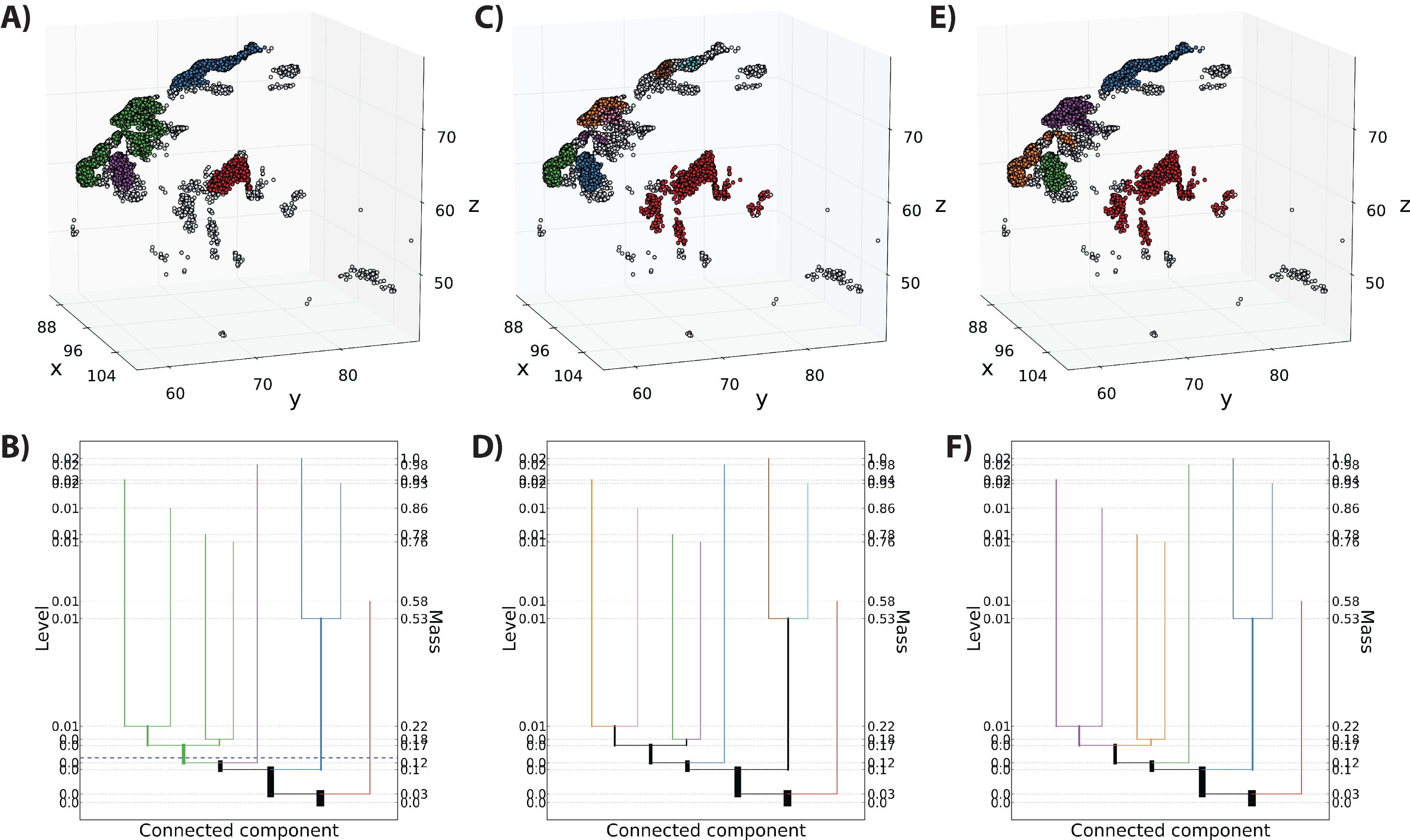}
	\end{center}
	\caption{{\bf Clustering with a level set tree.} A, C, E) Striatal endpoints
	colored by cluster assignment for three different cluster labeling methods.
	Gray points are unassigned because their estimated density is too low.
	Cluster colors match the tree node colors in the panels below. B) Tree nodes
	corresponding to clusters in panel A. These nodes are selected by cutting
	across the tree at a desired density or mass level. This method is very
	straightforward because it is based on the definition of the tree.
	D) Tree nodes corresponding to clusters in panel C. Each leaf of the tree
	produces a cluster. F) Tree nodes corresponding to clusters in panel E.
	The tree is traversed upward from the root (or roots) until the desired
	number of clusters first appears.}
	\label{fig:figure4}
\end{figure}

By construction, the tree is a compilation of connected components over the
levels of a pdf estimate, so the most straightforward way to cluster is to
retrieve the connected components at a chosen density or mass level (Figures 4A
and 4B). In addition to its definitional nature, this method conveys the most
intuitive sense for where the highest density data subsets are located. It also
allows the investigator to control the number of points in the clusters;
choosing a low mass level produces clusters that contain most of the data, while
high mass thresholds produce clusters with only the peaks of the data modes.
Finally, this method avoids the need to specify \textit{a priori} the number of
clusters, which must be chosen heuristically in many popular clustering methods
(k-means, for example).

The drawback of clustering at a single level is that it requires an arbitrary
choice of $\lambda$ or $\alpha$. All-mode clustering, which uses each leaf node
of a level set tree as a cluster, also automatically chooses the number of
clusters and avoids the arbitrary choice of a density or mass level at which to
cut the tree~\cite{Azzalini2007} (Figures 4C and 4D). This method does remain
sensitive to the choice of smoothing and pruning parameters, however. For a
given degree of pruning, this method tends to produce more and smaller clusters
than level set clustering.

If the clustering task demands a pre-set number of clusters, $K$, this can be
done with a level set tree by identifying the first $K$ disjoint components to
appear in the tree as the level increases from $\lambda=0$. Unlike k-means (and
related methods), there is no guarantee that there will be $K$ disjoint nodes in
a level set tree (Figures 4E and 4F).

In general, each method of labeling can capture general streamline clusters
approximately near macroscopic divisions of the striatal nuclei. For instance,
the red branch in each panel of Figure 4 highlights an isolated cluster of
prefrontal projections that terminate on the putamen.  The remaining clusters on
the caudate nucleus also break down into two major sets of endpoints. One set
(dark blue in Figures 4A and 4E, brown and cyan in Figure 4C) identifies
clusters of streamlines that terminate on the tail of the caudate, while the
third major set (green and purple in Figure 4A; orange, green and blue in Figure
4C; orange, green and purple in Figure 4E) identifies streamlines terminating
about the shell of the caudate nucleus. Thus, the first three branches of the
level set tree appear to capture known anatomical sub-divisions of inputs to the
striatum, with slight differences in sub-cluster identification depending on the
labeling approach used.

Each of these three methods typically assigns cluster labels to a fraction of
the sample, which we call the foreground points. The by-product of this is the
intelligent removal of outliers. Figure 4 shows that the
size of the foreground and outlier sets can vary greatly depending on the choice
of clustering method and parameter values. In particular, the all-modes
technique tends to create a large number of small clusters. When a full
segmentation is needed, the unlabeled background points can be assigned to a
cluster with any classification technique.

Together, the advantages of a level set tree approach---avoiding the need to
specify the cluster number, multiple clustering methods, visualization of many
cluster permutations, interactive clustering exploration, and automatic outlier
removal---allow the practitioner to gain greater insight into the clustering
behavior of a data set, using fewer assumptions than would be necessary for
standard methods.

\subsection*{Clustering performance evaluation}

To analyze the effectiveness of level set tree clustering we tested it in a
range of simulations against several standard clustering methods: k-means++,
gaussian mixture models, hierarchical agglomeration, spectral clustering,
diffusion map clustering, and DBSCAN. The simulations ranged in difficulty over
both the degree of separation of the clusters and the type of data generating
process, with the most complex scenario closely mimicking fiber track endpoint
distributions (see Methods for more detail).
\begin{figure}[!ht]
	\begin{center}
		\includegraphics[width=0.65\textwidth]{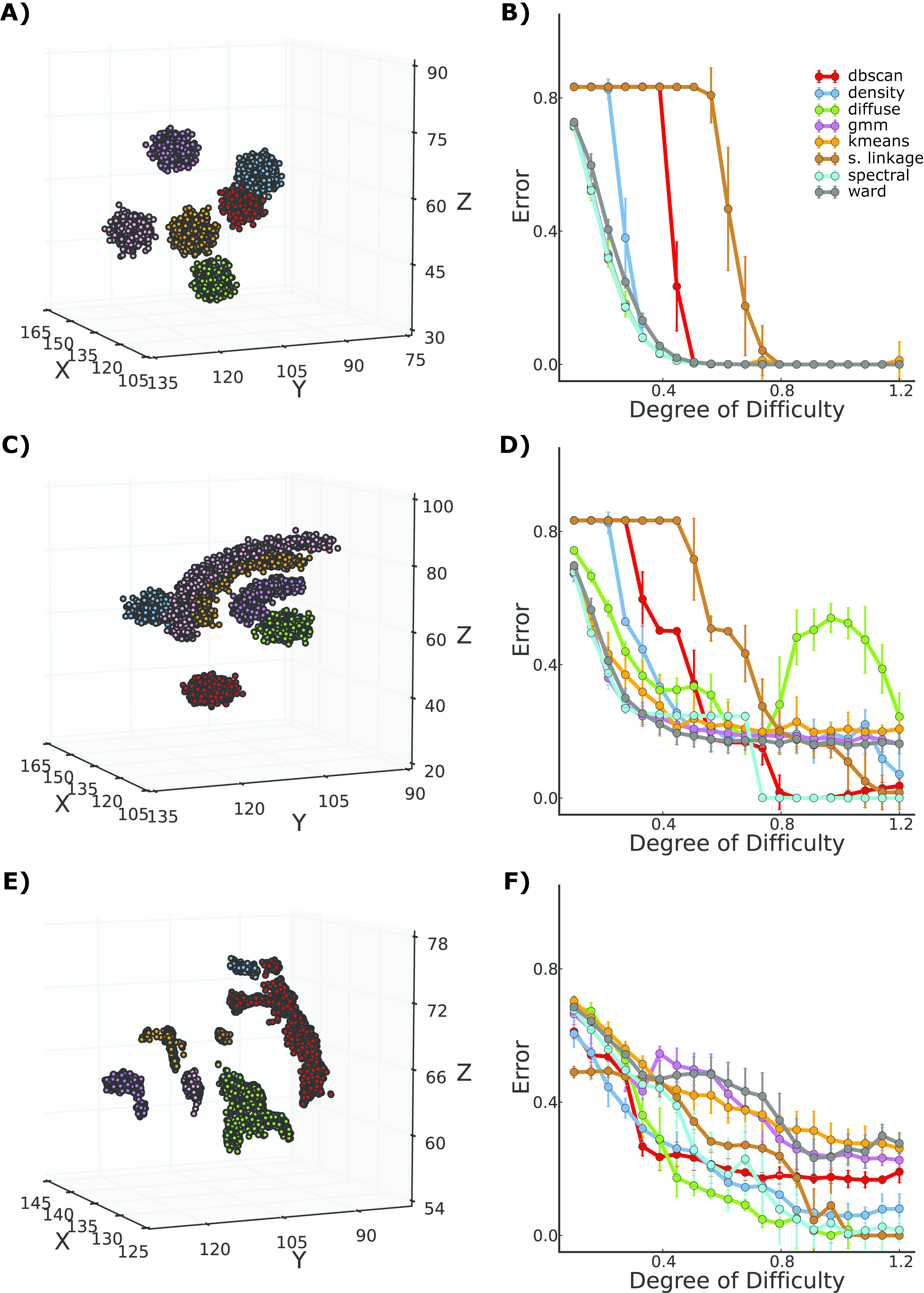}
	\end{center}
	\caption{{\bf Comparison of clustering method accuracy in simulations.} A,
	C, E) Example draws from each of three simulation scenarios (Gaussians, arcs
	\& Gaussians, and resampled striatal endpoints, from the top), with
	observations colored by true group label. B, D, E) Error rate for each type
	of simulation. For each simulation, the degree of clustering difficulty was
	varied by contracting the groups toward the grand mean by various amounts,
	so that most methods are accurate when the groups are well-separated and
	inaccurate when the groups overlap. For each type of simulation and each
	degree of difficulty, the mean and standard deviation of classification
	error are reported for 8 clustering methods: DBSCAN (dbscan), level set tree
	clustering (density), diffusion maps (diffuse), Gaussian mixture models
	(gmm), k-means++ (kmeans), hierarchical clustering with single linkage
	(s.link), spectral clustering (spectral), and hierarchical clustering with
	linkage by the Ward criterion (ward).}
	\label{fig:figure5}
\end{figure}

Not surprisingly, for the easiest clustering task---a mixture of six spherical
Gaussian distributions---all methods achieved perfect identification of the true
clusters when the groups were well separated. Single linkage hierarchical
clustering had a very high error rate even at medium degrees of separation
between clusters, due to the well-studied problem of chaining. The density-based
methods DBSCAN and level set trees also required more separation between
clusters before achieving the same error rate as parametric methods, possibly
due to the challenge of assigning low-density points to a cluster.

The results are more difficult to interpret for the moderately difficult
simulation scenario with three arcs and three spherical Gaussians. Single
linkage hierarchical clustering again required the most separation between
clusters to achieve highly accurate classification. Spectral clustering was
perfect when the clusters were well-separated and was as good as any other
method when the clusters were very close, but performed poorly at mid-range
degrees of separation. The closely related technique of diffusion maps actually
became less accurate at large degrees of separation. Level set tree clustering
performed poorly for tightly packed clusters, but was comparable to the
parametric methods (K-means++, Ward linkage, and GMM) for somewhat and very well
separated clusters.

While the second simulation type is more challenging than the mixture of only
Gaussians, it remains much simpler than the highly non-convex, multi-resolution
data in many fiber tractography clustering tasks. To test the clustering methods
in a more realistic setting we generated simulated data sets by randomly
resampling 5,000 points from a set of 10,000 striatal white matter fiber track
endpoints from one subject and adding Gaussian noise to each resampled point.
True group labels were assigned with a careful application of level set
clustering, and as with the the easy and moderate scenarios, we modulated the
difficulty of the task by contracting the clusters toward the grand mean to
varying degrees.

In this much more complicated and realistic setting, the parametric methods
performed poorly, achieving only about 70\% accuracy, even when the clusters are
very well separated. Each of the nonparametric methods (level set clustering,
DBSCAN, diffusion maps, and spectral clustering) performed best at some degree
of separation, making it difficult to identify clearly superior or inferior
methods. DBSCAN and level set trees have accuracies somewhat less than 100\%
even for well-separated clusters, probably due to the problem of assigning
low-density points to clusters. A more nuanced classifier for this step in level
set tree clustering would likely improve the results for level set trees in
particular.

Level set trees enjoy several categorical advantages over methods like spectral
clustering and diffusion maps, namely a more intuitive representation of data
structure, facilitation of interactive data exploration, a concise represention
of many different clustering permutations, and automatic selection of the number
of clusters. This experiment suggests level set trees are also at least as
accurate in practical clustering tasks, particularly with challenging nonconvex
clusters.

Finally, we take note that we have made no attempt to choose the parameter $k$
in an optimal manner in our experiments. 

\subsection*{Whole fiber segmentation}
\label{ssec:fiber-segment}
So far our analysis has focused on level set trees that are generated using just
the 3-dimensional endpoint locations of the fiber streamlines that terminate in
the striatum. This ignores the rich data contained in the rest of each fiber
streamline, which can provide substantially more information about differences
between sets of fibers than just the location of the streamline ends. We adopted
a pseudo-density approach for whole-fiber level set trees, where a pairwise
fiber distance is used to rank each streamline according to the spatial
proximity of its neighbors (see Methods). This pseudo-density ranking is not a
true pdf because it is not normalized to sum to one, but it can be used just
like a density function to define and index upper level sets. We use the
max-average-min fiber distance, which matches each point on a fiber to the
closest point on a second fiber, then averages these distance and finds the
maximum of the averages over the fiber pair (see Methods). In this way the level
set tree uses non-Euclidean distance as a factor in the whole-fiber clustering
process.

We used the pseudo-density level set tree approach to look at the organization
of cortico-striatal projections from two areas, the lateral frontal cortex and
orbitofrontal cortex, in the 30 subject template brain (Figure 6). In the
lateral frontal cortex we detected seven clusters of streamlines (34,982
foreground fibers out of 51,126 total fibers) that were organized in a
consistent, evenly spaced rostral-caudal direction along the middle frontal
gyrus (Figure 6A), an organization that is consistent with previous reports in
both the animal and human literatures~\cite{Draganski2008, Haber2010,
Verstynen2012}. Each identified cluster reflects regions of high pseudo-density
along the middle frontal gyrus. It is important to note that this whole-fiber
clustering was able to capture divergent patterns in the white matter pathways.
The dark blue and cyan streamlines start in the same region of the middle
frontal gyrus, but diverge to different sub-cortical targets (namely, the
caudate and putamen). This split is easy to identify in the level set tree by
the emergence of an early branching in the tree into two major divisions that
reflect caudate versus putamen fibers (Figure 5B). This provides a clean
anatomical segmentation of the fibers despite the fact that these two fiber sets
start in the same region of the middle frontal gyrus.
\begin{figure}[!ht]
	\begin{center}
		\includegraphics[width=0.95\textwidth]{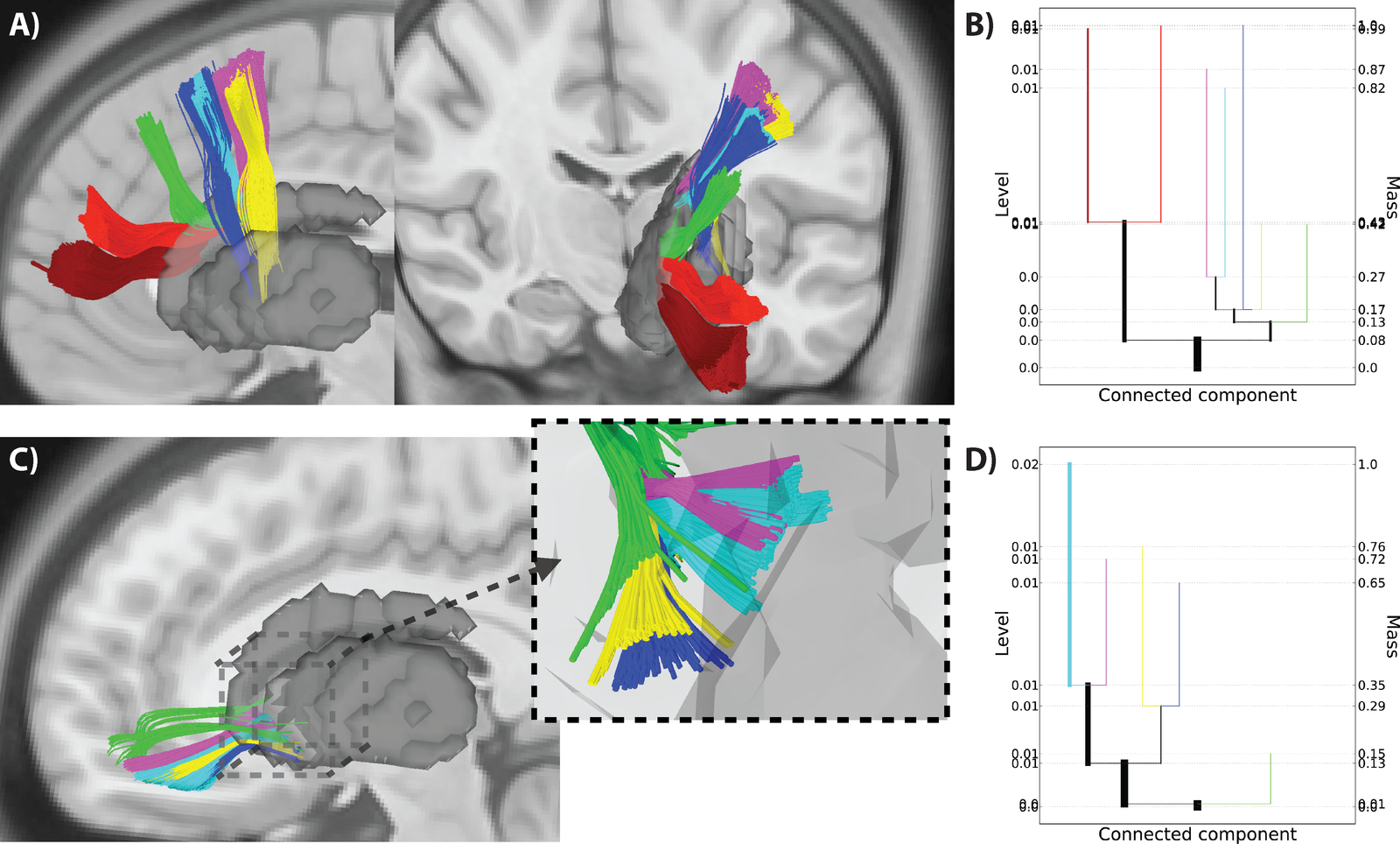}
	\end{center}
	\caption{{\bf Level set tree clustering for whole fiber streamlines.} A)
	Foreground fibers for the seven selected clusters from the 30 subject
	template data set for streamlines tracked between the middle frontal gyrus
	and striatum, shown in both a sagittal and coronal view. Clusters are
	colored according to an all-mode clustering of the tree. B) The level set
	tree for data in panel A. Tree leaves are matched to fiber clusters by
	color. C) Same analysis as shown in A, but for a set of streamlines from the
	orbitofrontal cortex. Inset shows closeup of fiber streamlines in the
	striatal ROI mask. D) Level set tree for data shown in panel C. The branch
	colors of trees in panels B and D match the clusters shown in the
	streamlines of panels A and C respectively.}
	\label{fig:figure6}
\end{figure}

In the projections from the orbitofrontal cortex we identified five mode
clusters (Figures 6C and 6D). Close inspection of the endpoints of these
streamlines in the striatum reveals that each cluster forms a striated-like
pattern in the caudate that is similar to patterns previously reported in
corticostriatal projections ~\cite{Verstynen2012} (Figure 6C, inset). These
striated formations are thought to reflect the modularized biochemical makeup of
the striatum~\cite{Graybiel1978, Ragsdale1990}. This complex arrangement is
difficult to capture with clustering methods that assume convex cluster shapes,
but the whole-fiber pseudo-density clustering approach successfully extracts the
patterns with minimal assumptions.

\subsection*{Level set tree variability}
\label{ssec:lst-stability}
To assess the stability of the 30 subject template level set trees in Figure 6,
we created a set of trees by subsampling from the original lateral and
orbitofrontal fiber streamline data sets and constructing a tree for each
subsample. This simulates the variability seen when repeating the tractography
on the same data set multiple times. The overlaid tree plots in Figures 7A and
7D indicate a high degree of stability for the trees built from these subsampled
data sets, although it appears that the stability might be slightly lower for
the orbitofrontal set. This conclusion is supported for both ROIs by the mode
function overlays and histograms of mass values where each tree splits. These
plots illustrate that the existence of each tree branch is consistent across the
subsamples, even though there is some variation in the mass levels where the
branches first appear.
\begin{figure}[!ht]
	\begin{center}
		\includegraphics[width=0.95\textwidth]{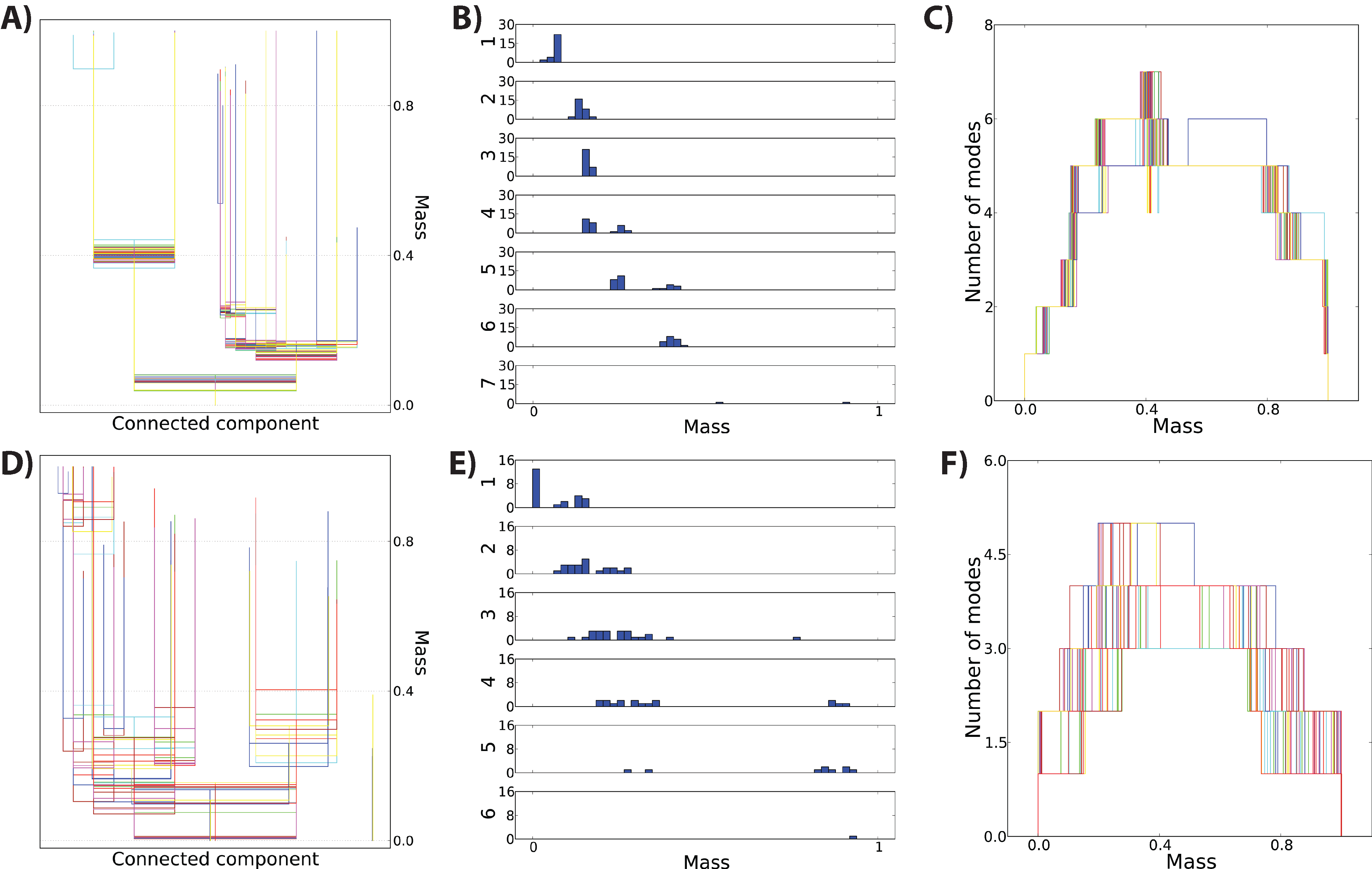}
	\end{center}
	\caption{{\bf Repeat reliability for level set tree results for a 30-subject
	template, using subsampling.} For the middle frontal gyrus ROI, 28 random
	subsamples of 15,000 fibers were drawn from the total of 51,126 fibers,
	while 1,500 fibers were drawn for 23 subsamples from the 3,038 total fibers
	in the rectus. A) All 28 level set trees plotted on the same canvas,
	illustrating the high degree of similarity between the data structure in the
	subsamples. B) Histograms of the mass levels of the splits over the whole
	set of subsample trees. Slit mass levels are matched across subsamples by
	rank order. C) All 28 mode functions plotted together, illustrating that
	there is little variation in the number of clusters at each mass level. D)
	All 23 level set trees plotted together. E) Distribution of mass values for
	splits, matched across subsamples by rank within each sample's tree. F) All
	23 mode functions overlaid.}
	\label{fig:figure7}
\end{figure}

The high degree of stability in these subsample trees conveys certainty to the
features of the level set trees constructed on the full data set (Figures 6B and
6D). For example, the left branch of the tree for the lateral frontal
projections contains two prominent nodes (red and dark red) that appear when 42
percent of the fibers are in the background (i.e.~not in the the upper level
set). The fact that this same split occurs in every one of the subsample trees
is evidence that such a split exists in true underlying (and unobserved)
distribution of fibers that generated this data set.

On the other hand, data sets that differ even in seemingly small ways can lead
to much more variation in the resulting level set trees. For a subset of
subjects, fiber streamlines were reconstructed for two separate scans separated
by six months. Figure 8 shows the level set trees constructed for the lateral
frontal projections from each scan in several example subjects, as well as the
foreground clusters produced by all-mode clustering. The foreground clusters
reveal that there does tend to be an overall high degree of similarity between
the fiber streamline sets across trials, with the exception of one or two
well-defined clusters that only appear in one of the two scans (highlighted in
gray in Figure 8).
\begin{figure}[!ht]
	\begin{center}
		\includegraphics[width=0.95\textwidth]{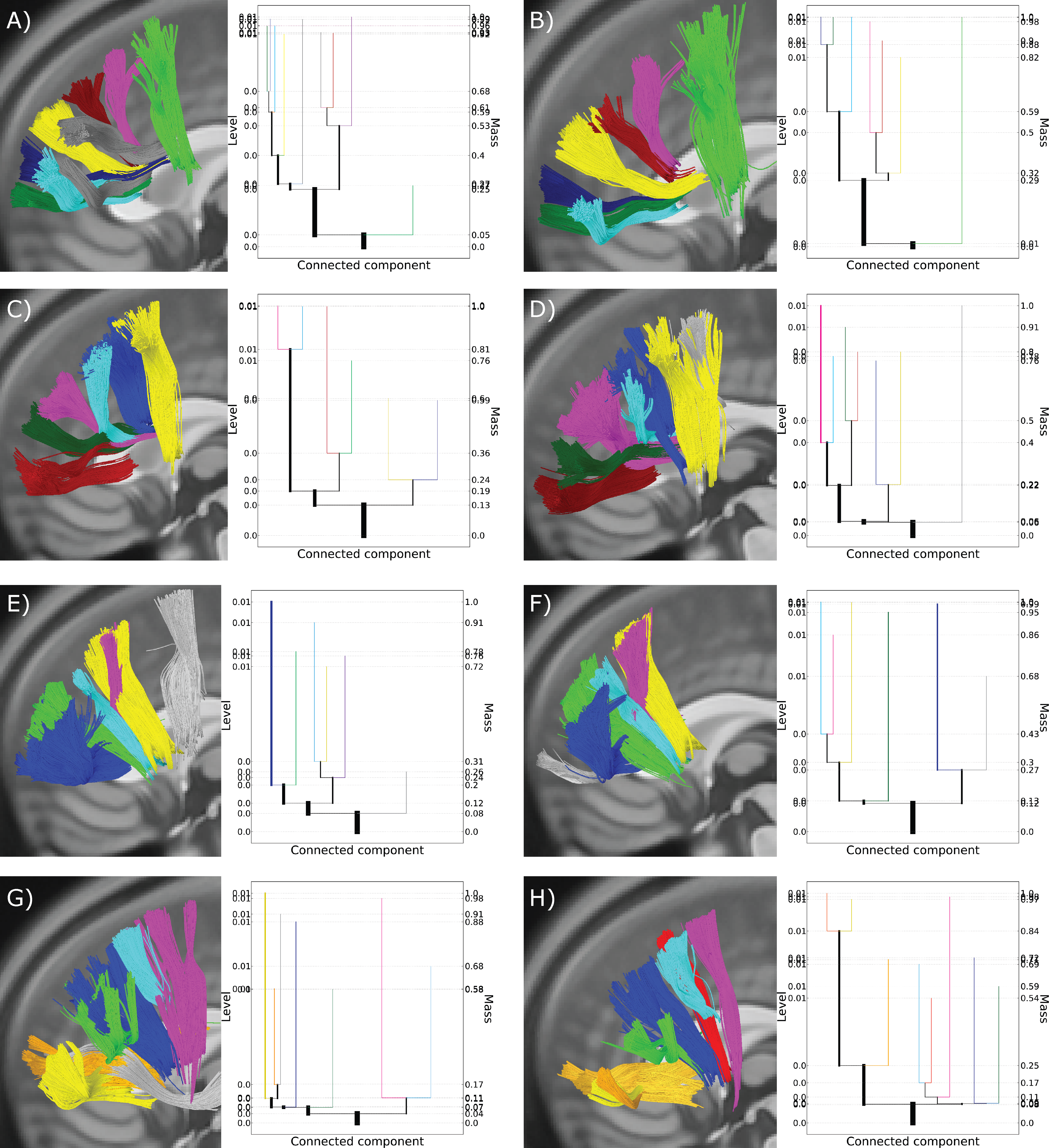}
	\end{center}
	\caption{{\bf Test-retest comparisons for four subjects, tested six months
	apart.} Colored streamlines show clusters that were consistently observed at
	both scan times. Gray streamlines show clusters detected at only one time
	point. Panels A, C, E, and G show results from the initial scan session.
	Panels B, D, F, and H show results from the second scanning session six
	months later.}
	\label{fig:figure8}
\end{figure}

The level set trees likewise reflect similar structure across scans, but the
non-overlapping tree nodes appear to exaggerate the differences between trees.
For example, panels E and F in Figure 8 show the foreground fiber streamlines
and level set trees for two scans of a single subject. The blue, green, cyan,
violet, and yellow clusters match well across scans and appear to share very
similar topography. However, panel E contains an obvious cluster on the right
side of the plot (in gray) that is not present in panel F, while panel F
contains its own obvious cluster on the left side of the plot (also in gray)
that is not present in panel E. Note that each branch's (or cluster's) color was
manually defined to match between images of the same subject, but does not
necessarily reflect the same branch/cluster identified across subjects. While
some of the features of the trees reflect the overall similarity---for example,
the number of leaves is the same and the yellow, cyan, and violet clusters are
more similar to each other while the blue cluster is much different---the
overall shape of the trees is very different.

These variations reflect actual differences between the test and retest data,
not just variability of the level set tree procedure. Not only are some clusters
present in only one of the two data sets (shown in gray in Figure 8), but
differing tree shapes and branching locations indicate that the probability
content and relative hierarchy of even similar-looking clusters is not the same
across scans. Despite such marked differences in the test and retest data sets,
the output from all-mode clustering retains a very high degree of consistency
across scan sessions, demonstrating the robustness of the proposed methodology.

\section*{Discussion}
\label{sec:disc}

White matter pathways have highly complex shapes and spatial organization,
making it difficult to summarize their topographic structure. We have shown that
level set trees provide a concise representation of this topography by
describing the hierarchy of modal regions in the density or pseudo-density
function that describes the probabilistic spatial distribution of a set of fiber
tracks. We demonstrated that this hierarchy is useful by identifying not only
major anatomical boundaries in the striatum (e.g., putamen vs. caudate), but
also sub-regions within the same nucleus (e.g., shell vs. tail of the caudate;
see Figure 4). We demonstrated the reliability of these results through
qualitative comparisons of level set trees in repeated sub-sampling and
test-retest experiments, suggesting that level set trees have the potential be
used as statistical estimators of fiber streamline topography. Finally, we
evaluted the performance of level set trees in several simulations against a
suite of standard clustering methods that are commonly used to describe fiber
streamline organizational patterns. Level set trees performed as well as any of
the clustering methods, however we should also emphasize that describing fiber
track topography and clustering these data points are not equivalent tasks. For
the purpose of summarizing topography, level set trees have several advantages
over traditional clustering techniques: they are statistically principled; they
are compact data structures that enable fast retrieval of high-density clusters
at any density level; they allow a multi-scale visualization of the cluster
patterns in a data set; they are a natural platform for interactive data
exploration; and they offer several methods for obtaining particular cluster
labels without assuming the number of clusters and with automatic removal of
outliers.

Level set trees are traditionally based on an estimate of an unobserved pdf that
is assumed to have generated a data set, which is a realistic assumption for
fiber streamline endpoints. We show for this type of data how level set trees
can be used to visualize data patterns, interactively explore structurally
coherent data subsets, and simultaneously present many different cluster
labelings. Where the assumption of a pdf is not realistic, as with
infinite-dimensional whole fibers, we extend level set trees to address the
problem of describing the topography by observing that a pseudo-density
estimator (along with a similarity measure) is sufficient for level set tree
construction.

Ideally, level set trees could be used as statistical tool for inference when
comparing white matter topographies across populations.  For example, does the
organization of fiber streamlines between two brain areas differ in individuals
with neurological disorders (e.g., autism) when compared to neurologically
healthy controls?  By qualitatively demonstrating the reliability of level set
tree structure, we highlight this potential of the method. Quantification of the
uncertainty in level set trees is an open research problem; the qualitative
comparisons shown in this paper as well as other preliminary work in this
direction~\cite{Ben-Hur2002, Rinaldo2012, Smith1980} have shown that level set
trees constructed on data drawn from the same distribution tend to be very
similar, or stable, while trees constructed on data drawn from different
distributions tend to be different. Stability is a difficult principle to apply,
however, because simply identifying when two trees are similar is also an open
research area (one that deserves future attention, but is well beyond the scope
of the current project).

An important limitation of our methodology is the selection of tuning parameters
$k$ for connectivity and density (or pseudo-density) estimation and $\gamma$ for
tree pruning (although the choice of cluster number is not required as with most
clustering methods). We could choose these parameters based on the optimal
values found in the theoretical literature~\cite{Rao1983} but these values are
only valid in asymptotic regimes (where the sample size increases) and tend to
work poorly in practice. As a result, as with much of applied statistics,
selecting tuning parameters requires sound empirical judgment. It should be
noted, however, that in our experiments the results tend to be robust for a
relatively large range of tuning parameter values. Because there are several
ways to obtain clusters from level set trees, inserting level set tree methods
into an automated data analysis pipeline also requires a choice of cluster
labeling method, in addition to the tuning parameters.

Despite these limitations, level set trees are a novel and powerful way to
analyze the topography of fiber streamline data sets with minimal \textit{a
priori} assumptions. As DWI methodologies improve, the usefulness of this
approach for characterizing sub-divisions in anatomical pathways will allow for
greater specificity of regions of interest. Originally intended to describe
probability density functions, level set trees can be extended to model
pseudo-density functions as well, allowing us to apply the trees' powerful data
visualization and clustering tools to the analysis of fiber streamline data
sets. This flexibility opens the potential for this method of density-based
clustering approaches to be used in a variety of neuroimaging analyses beyond
white matter tractography. Future work will focus on these extended applications
in a neuroimaging context.

\bibliography{biblio/densclust,biblio/neuro,biblio/dsiclust,biblio/functional,biblio/books,biblio/genclust}

\end{document}